\newcommand{\be}{\begin{equation}}
\newcommand{\ee}{\end{equation}}
\newcommand{\bea}{\begin{eqnarray}}
\newcommand{\eea}{\end{eqnarray}}
\newcommand{\ba}[1]{\begin{array}{#1}}
\newcommand{\ea}{\end{array}}
\documentclass[twocolumn,secnumarabic,amssymb,nobibnotes,nofootinbib,aps,pra,showpacs]{revtex4}
\usepackage{epsfig}
\usepackage{amssymb}
\begin{document}
\setlength{\topmargin}{-0.05in}

\title{Ion-atom cold collision: Formation of cold molecular ion by radiative processes}
\author{Arpita Rakshit$^1$ and Bimalendu Deb$^{1,2}$}
\affiliation{$^1$Department of Materials Science, and 
$^2$Raman Center for Atomic, Molecular and Optical Sciences, Indian Association
for the Cultivation of Science,
Jadavpur, Kolkata 700032, India.}

%\date{\today}

\def\zbf#1{{\bf {#1}}}
\def\bfm#1{\mbox{\boldmath $#1$}}
\def\hf{\frac{1}{2}}
\begin{abstract}
We discuss theoretically ion-atom collisions at low energy and predict the possibility of formation of cold molecular ion by photoassociation. We present results on radiative  homo- and hetero-nuclear atom-ion cold collisions that reveal threshold behaviour of atom-ion systems.
\end{abstract}

\maketitle

\section{introduction}
\begin{figure}
\includegraphics[width=\columnwidth]{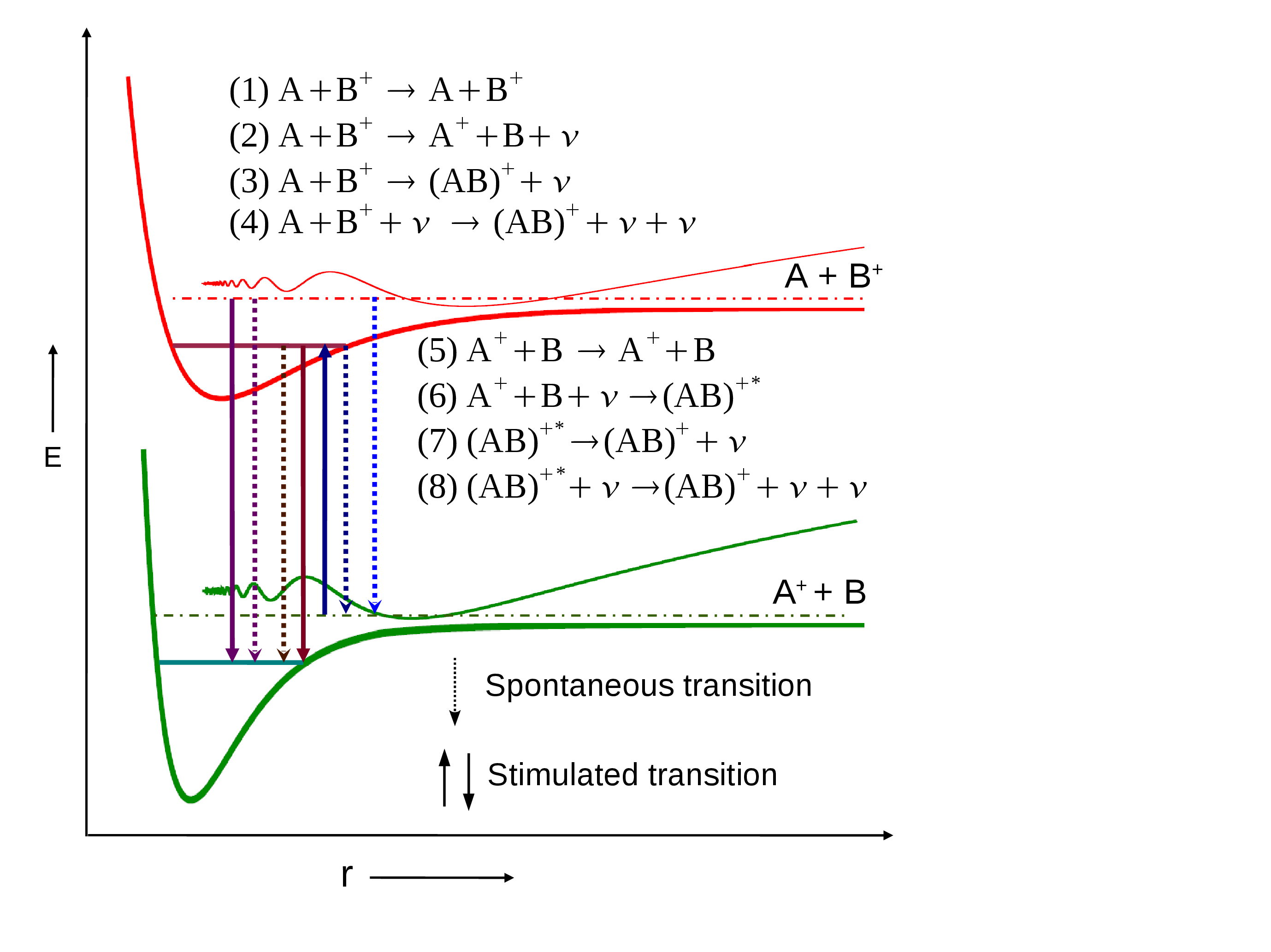}
\caption{(Color online) Schematic diagram of  possible physical processes which can take place during atom-ion collision at low energy. }
\end{figure}

Molecular ions are important for a variety of fundamental studies in physics. For instance, it is proposed that cold molecular ions would be useful for measuring electron dipole moment (EDM) \cite{cornell,meyer}.  Study of cold molecular ions has relevance in diverse areas such as metrology \cite{metro1,metro2} and astrochemistry \cite{astrochem}.  Recently, molecular ions are cooled into ro-vibrational ground states  by   all optical  \cite{Np6-275}, laser and sympathetic cooling methods \cite{PRA79-032716, Np6-271}. A large variety of diatomic and triatomic molecular ions are also cooled by sympathetic method \cite{Roth, Ostendorf, Molhave}. Other methods such as photoassociative ionisation \cite{Bag, PRL60-788, PRL70-2074, PRL70-3225, Sch}, buffer gas \cite{Pearson}, and rotational cooling  \cite{Vogelius}  have been widely used for producing low energy molecular ions. Since cooling of neutral atoms and atomic ions down to sub-milliKelvin temperature regime is possible with currently available technology of laser cooling, it is now natural to ask ourselves: Is it possible to form cold molecular ion by atom-ion cold collision? Recent progress in developing hybrid traps \cite{hytrap, nl464-388, arXiv1005:3846v1,PRL102-223201} where both atomic ions and neutral atoms can be simultaneously confined provides new opportunity for exploring ion-atom quantum dynamics and charge transfer reactions at ultralow temperatures.  As neutral cold atoms can be photoassociated \cite{PA} into cold dimers, the same association method should also apply to atoms colliding with atomic ions forming cold molecular ions.  
%To address the issues of ion-atom Photoassociation (PA) and Stimulated Radiative Charge Transfer (SRCT), it is first necessary to %understand the interaction between an atom and an ion at low energy in the presence of an external electromagnetic field.
\begin{figure}
\includegraphics[height=3.25in, width=\columnwidth]{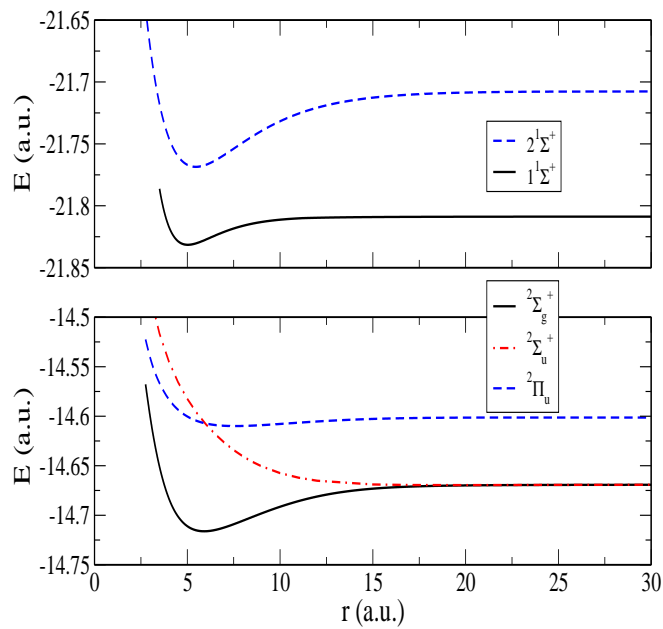}
\caption{(Color online) Upper panel shows $1^1\Sigma^+$ (solid) and $2^1\Sigma^+$ (dashed) model  potentials of (LiBe)$^+$ system.  Lower panel shows  $^2\Sigma_g^+$ (solid)), $^2\Sigma_u^+$ (dashed-dotted) and $^2\Pi_u$ (dashed)  potentials of Li$_2^+$.}
\end{figure}
Understanding ion-atom cold collision \cite{PRL102-223201, nl464-388, arXiv1005:3846v1, Schmid,  NJP10- 033204, Zhang, yb, PRA79-010702(R)2009, BoGao, PRA62-012709,Marakov,Ma} is important for realizing a charged quantum gas, studying charge transport \cite{transport} at low temperature, exploring polaron physics \cite{polaron1,polaron2,polaron3} and producing ion-atom bound-states \cite{bound} and  cold molecular ions \cite{Np6-271,PRA79-032716,Np6-275}. 

Although in recent times there have been several studies on ion-atom cold collisions, formation of molecular ion by photoassociation (PA) is yet to be demonstrated. There are qualitative differences between atom-atom and ion-atom PA. In contrast to atom-atom PA, hetero-nuclear atom-ion PA is accompanied by charge transfer. Neutral atom-atom PA involves excited diatomic molecular states which in the separated-atom limit  correspond  of one ground (S) atom and the other excited (P) atom.  Hetero-nuclear atom-ion PA may involve excited molecular states which asymptotically correspond to separated atom and ion both belonging to S electronic states. The long-range potentials of ion-atom system behave quite differently from those of neutral atom-atom system.  

Here we show that it is possible to form translationally and rotationally cold  molecular ion by  PA. We specifically focus on hetero-nuclear radiative processes. However, we study in general both homo- and hetero-nuclear ion-atom cold collisions to reveal the contrast between the two processes. At ultralow collision energies, radiative charge transfer processes dominate over non-radiative ones. Starting from  a cold alkaline metal earth ion and an ultracold alkali atom (such as an atom of alkali Bose-Einstein condensates) as the initial reactants, formation of ground state molecular ion requires a three-step radiative reaction process. In the first step, the ion-atom pair in the continuum of the excited electronic state undergoes radiative charge transfer to the continuum of the ground electronic state. In the second step, the ground continuum ion-atom pair is exposed to laser radiation of appropriate frequency to photoassociate them into excited molecular ion. In the third and final step, another laser is used to stimulate the excited molecular ion to deexcite into a particular rovibrational level of ground electronic state. Since molecular ion is formed from initially cold atom and ion, the molecular ion remains translatioanlly and rotationally cold. One noteworthy feature of this method is the selectivity of low lying rotational level.   We present selective results on  elastic and radiative charge transfer scattering cross sections for both homo- and hetero-nuclear ion-atom collisions.  For model potentials of (LiBe)$^+$ system, we calculate the PA rate of formation of LiBe$^+$ molecular ion.
\begin{figure}
\includegraphics[width=\columnwidth]{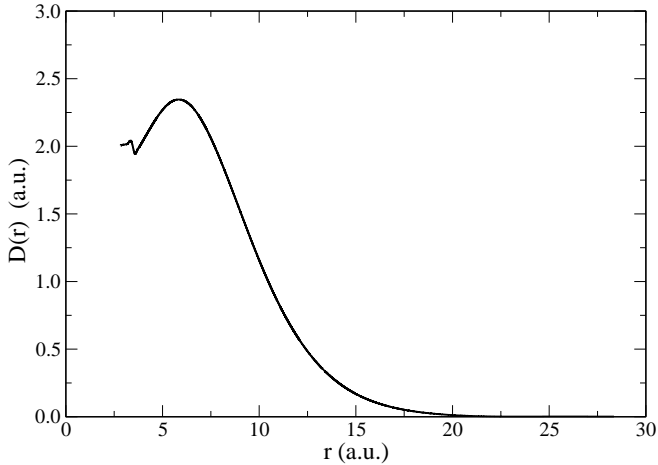}
\caption{Radial transition dipole matrix element as a function of separation $r$ for (LiBe)$^+$ system.}
\end{figure} 
\begin{figure}
\includegraphics[width=\columnwidth]{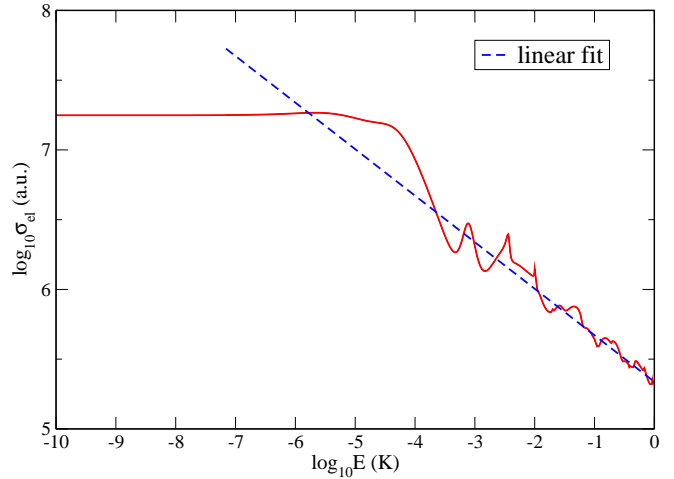}
\caption{(Color online) Total elastic scattering cross-section $\sigma_{el}$ for Li + Be$^+$ ($2^1\Sigma^+$) collision is plotted against collision energy $E$ in K. The dashed curve is a linear fit for energies greater than 10$^{-6}$ K.}
\end{figure}

This paper is organised in the following way. In Sec.2, we describe our model focussing on possible elastic and inelastic processes. Results are presented and discussed in Sec.3. In the last section we draw our conclusions. 
 %=======================================================================================================
%\maketitle

\section{Elastic and Inelastic Processes}
\begin{table}
\caption{Dissociation energies $D_e$ in a.u., equilibrium positions $r_e$  and  the effective lengths $\beta_4$ in Bohr radius for excited and ground state potentials ($V(r)$) of  (LiBe)$^+$ and (LiLi)$^+$ systems.}
\begin{tabular}{c  c  c  c  c  c  c c c }
\hline 
\multicolumn{1}{c}{{system}} & \multicolumn{1}{c}{\vline} & \multicolumn{1}{c} {V(r)} &  \multicolumn{1}{c}{\vline} & \multicolumn{1}{c} {$D_e$} &  \multicolumn{1}{c}{\vline}  & \multicolumn{1}{c} {$r_e$} &  \multicolumn{1}{c}{\vline}  & \multicolumn{1}{c} {$\beta_4$}  \\
\hline
(LiBe)$^+$ & \vline & $2^1\Sigma^+$ &\vline & 0.06 & \vline &  5.46 & \vline & 1083.4  \\
(LiLi)$^+$ & \vline & $^2\Pi_u$ & \vline  & 0.01  &  \vline &  7.50 &  \vline &  1019.8\\
(LiBe)$^+$ & \vline & $1^1\Sigma^+$ & \vline  & 0.02 &  \vline &  5.03 &  \vline & 515.9 \\
(LiLi)$^+$ & \vline & $^2\Sigma_g^+$ & \vline  & 0.05  &  \vline & 6.00 &  \vline & 1019.8\\  
\hline
  \end{tabular}
\label{tb3}
\end{table} 
We consider cold collision of an alkali atom A with an alkaline earth metal ion B$^+$ in a hybrid trap. The possible elastic and inelastic processes are schematically depicted in Fig. 1. These are : (1) elastic collision between A  and B$^+$, (2) an e$^-$ from  A may hop to B$^+$ provided they are close enough to each other forming ground state pair of ion A$^+$  and atom B, (3) atom-ion pair in the excited continuum may decay spontaneously to a bound level of lower electronic state, (4) excited atom-ion pair may be transferred to a ground electronic bound state by stimulated emission process, (5) the ground state atom-ion pair may undergo elastic collision, (6) the ground pair may be photoassociated in the presence of appropriate laser  radiation to form excited molecular ion, (7) this excited molecular ion may decay sponataneously either to a ground bound state or  continuum, (8) excited bound state may be transferred to a ground bound state by stimulated emission process. 

\begin{figure}
\includegraphics[height=2.0in,width=\columnwidth]{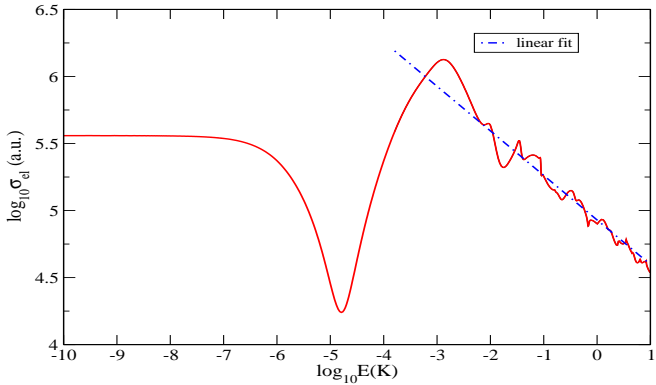}
\caption{(Color online)Same as in Fig.4 but for Li$^+ +$ Be ($1^1\Sigma^+$).}
\end{figure}

To illustrate atom-ion radiative cold collisions, we consider a model system of $^7$Li + Be$^+$ undergoing elastic and radiative charge transfer collisions. The possible experimental situation can be imagined as a single Be$^+$ ion immersed in a Bose-Einstein condensate of $^7$Li atoms in a hybrid trap. The molecular potentials $1^1\Sigma^+$  (ground) and $2^1\Sigma^+$  (excited) of (LiBe)$^+$ system as shown in Fig.2 (upper inset) asymptotically go to $^1$S $+ ^1$S (Li$^+ +$ Be) and $^2$S $+ ^2$S (Be$^+ +$ Li), respectively. We construct  model potentials $1^1\Sigma^+$  (ground) and $2^1\Sigma^+$  (excited) of (LiBe)$^+$ system using spectroscopic constants given in Ref.\cite{Safonov}. Short range potential is approximated using Morse potential and the long range potential \cite{PRA62-012709, Marakov} is given by the expression
\begin{equation}
V(r) = -\frac{1}{2}\left( \frac{C_4}{r^4} + \frac{C_6}{r^6} +  \cdots \right) 
\end{equation}
where $C_4$, $C_6$ correspond to dipole, quadrupole polarisabilities of atom concerned. The polarisation interaction falls off much more slowly than van der Waals interaction which represents the long range part of interaction between neutral atoms. Hence collision between atom and ion is dominated by the long range polarization interaction. The qualitative feature of this long range interaction of atom-ion is governed by effective length which is given by $\beta_4 = \sqrt{2\mu C_4/\hbar^2}$ where $\mu$ is the reduced mass.
The short range and long range parts of the potentials  are  smoothly joined by spline. 

Since Li$^+$ may be formed due to charge transfer collision between Be$^+$ and Li, we need to consider the interaction between this Li$^+$ and other Li atoms present in the condensate. The data for $^2\Sigma_g^+$ , $^2\Sigma_u^+$  and $^2\Pi_u$ potentials of Li$_2^+$ are taken from Ref.\cite{KONOWALOW}. Dissociation energy $D_e$, equilibrium position $r_e$ and effective range $\beta_4$ of the ground and excited state potentials of (LiBe)$^+$ and LiLi$^+$ systems are given in Table I. A comparison of potentials of these two systems reveals that  ground state potential $1^1\Sigma^+$ of (LiBe)$^+$ is much shallower than  $^2\Sigma_g^+$ potential of Li$_2^+$. The equilibrium positions of both ground and excited state potentials of (LiBe)$^+$ system lie almost at the same separation. Unlike the asymptotic behavior of the excited $2^1\Sigma^+$ potential of (LiBe)$^+$ system, the excited state potential $^2\Pi_u$ of homonuclear Li$_2^+$ molecular ion   asymptotically corresponds to one Li$^{+}$ ion in the electronic ground S state and one neutral Li atom in the excited P state.  The equilibrium positions $r_e$ of ground and excited state potentials of Li$_2^+$ system are shifted by  1.5 Bohr radius.  For of (LiBe)$^+$ system, we notice that $\beta_4$ of excited ($2^1\Sigma^+$) potential is almost twice that of the ground ($1^1\Sigma^+$) potential.

\begin{figure}
\includegraphics[height=2.0in,width=\columnwidth]{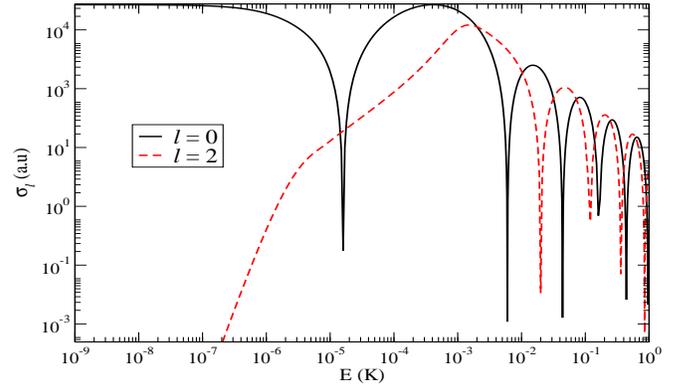}
\caption{(Color online)Partial wave cross sections for Li$^+ +$ Be ($1^1\Sigma^+$) collision are plotted as a function of $E$ (in K) for $\ell = 0$ (solid) and $\ell = 2$ (dashed).}
\end{figure}

Let us first consider cold collision between Li and Be$^+$ with both of them being in $^2$S electronic state. So,  our initial system corresponds to the continuum of $2^1\Sigma^+$  potential. Due to charge transfer collision neutal Be atom and Li$^+$ ion are generated.  In the separated two-particle limit of this system, dipole transition to ground state at the single particle level is forbidden. Furthermore, since at low energy non-radiative charge transfer is suppressed, the dominant inelastic channel is the radiative charge transfer transition that occurs at intermediate or short separations. Electronic transition dipole moment between two ionic molecular electronic states vanishes at large separation. Therefore, transitions occur at short range where hyperfine interaction is negligible in comparison to central(Coulombic) interaction.  The total molecular angular momentum is given by $\vec{J} = \vec{S} + \vec{L} + \vec{\ell}$ where $S$ and $L$ are the total electronic spin and orbital quantum number, respectively;  and $\ell$ stands for the angular quantum number of the relative motion of the two atoms. For the particular model for (LiBe)$^+$ system chosen here, we have $L=0$ and $S=0$ for both the ground and the excited electronic states. Thus here the total angular momenta for both the ground and excited states  are given by  $J = \ell$. However, it is more appropriate to denote total angular quantum number of a molecular bound state by $J$ and that of the continuum or collisional state of this atom-ion  system by simply $\ell$. The parity selection rule for the electric dipole transition between the ground and excited states dictates $\Delta J = \pm 1$.

\begin{figure}
\includegraphics[height=2.5in,width=\columnwidth]{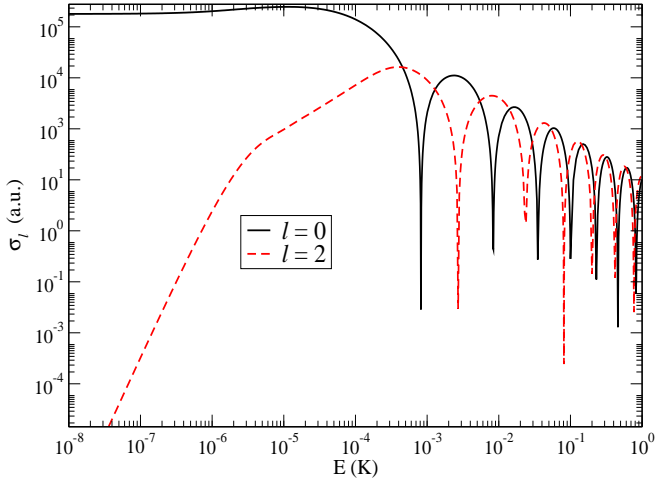}
\caption{(Color online) Same as in Fig. 6 but for Li $+$ Li$^+$ ground state collision in $^2\Sigma_g$ state.}
\end{figure}

To investigate ion-atom elestic scattering and free-bound transitions, we need to calculate continuum wave functions which are obtained by solving the partial wave Schr\"{o}dinger equation given by 
\bea
\left[ \frac{d^2}{dr^2} + k^2 - \frac{2\mu}{\hbar^2} V(r) - \frac{\ell (\ell +1)}{r^2} \right ] \psi_{\ell} (kr) = 0 
\eea
where $r$ is the ion-atom separation. The wave function $ \psi_{\ell}(kr) $ has the asymptotic form $ \psi_{\ell}(kr) \sim \sin\left[ kr - \ell\pi/2 + \eta_l\right] $ with $\eta_\ell (k)$ being the phase shift for $\ell$-th partial wave.  The total elastic scattering cross section is expressed as
\begin{equation}
\sigma_{el} = \frac{4\pi}{k^2}\sum_{\ell=0}^\infty (2\ell+1)\sin^2(\eta_\ell)
\end{equation}
where $ k = \sqrt{(2mE/\hbar^2)}$. As the energy gradually increases more and more partial waves start to contribute to total elastic scattering cross sections and the scattering cross section at large energy is  \cite{PRA62-012709}
\bea
\sigma_{el} \sim \pi \left( \frac{\mu C_4^2}{\hbar^2}\right)^\frac{1}{3} \left(1 + \frac{\pi^2}{16}\right) E^{-\frac{1}{3}}\label{sig}
\eea 
 As $k \rightarrow 0$, according to Wigner threshold laws  $\eta_{\ell}(k) \sim k^{2 \ell + 1}$  if $\ell \le (n-3)/2$ with $n$ being the exponent of long-range potential behaving as  $ \sim 1/r^n$ as $r \rightarrow \infty$. If  $\ell > (n-3)/2$ then the threshold law is  $\eta_{\ell}(k) \sim k^{n - 2}$. Since the long-range part of ground as well as excited ion-atom potentials  goes as $\sim 1/r^4$ as $r \rightarrow \infty$, Wigner threshold laws tell us that  s-wave ($\ell = 0$) ion-atom scattering cross section should be independent of $k$ while all the higher partial wave scattering cross sections should go as $\sim k^2$ in the limit $k \rightarrow 0$. 

\begin{figure}
\includegraphics[height=2.5in,width=\columnwidth]{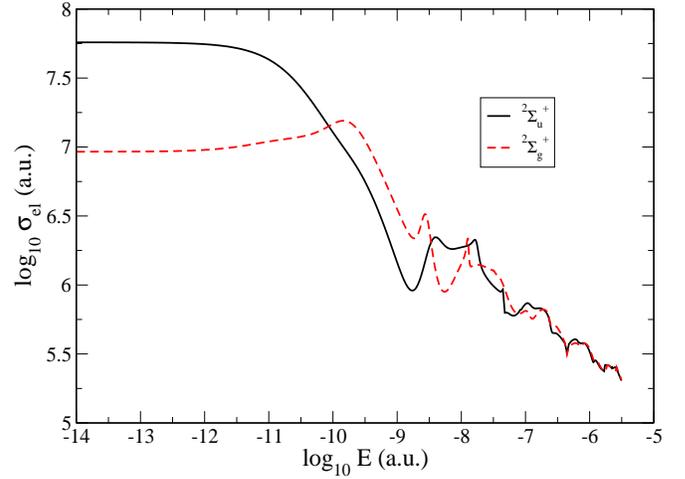}
\caption{(Color online)Same as in Fig.4 but for Li $+$ Li$^+$ collision in $^2\Sigma_g^+$ (dashed) and  $^2\Sigma_u^+$ (solid) potentials. }
\end{figure}
Ion-atom inelastic collisions are mainly of two kinds - charge transfer reactions and radiative- or photo-associative transfer \cite{Cooper,Zygelman1,Zygelman2,Stancil,West}.
The radiative charge transfer cross section \cite{Cooper,Zygelman1,Zygelman2} is given by 
\bea
\sigma_{ct} = \int_{\omega_{min}}^{\omega_{max}} \frac{d\sigma_{ct}}{d\omega} d\omega \label{ct}
\eea 
where $\omega$ is the angular frequency of emitted photon and
\bea
 \frac{d\sigma_{ct}}{d\omega} &=&  \frac{ 8\omega^3\pi^2}{3 c^3 k_m^2} 
\sum_l\left [ \ell M_{\ell,\ell-1}^2(k_m,k_n)  \right. \nonumber \\ &+& \left. (\ell+1)M_{\ell,\ell+1}^2(k_m,k_n)\right ] 
\eea 
where
\bea
  M_{\ell,\ell'}(k_m,k_n) = \int_0^\infty dr \psi_\ell^m(k_mr)D(r)\psi_{\ell'}^n(k_nr) 
\eea
$D(r)$ is the magnitude of the molecular transition dipole moment. Here $ k_m = \sqrt{2\mu\left[ E - V_m(\infty)\right] }$ and $ k_n = \sqrt{2\mu\left[ E - V_n(\infty) - \hbar\omega \right] }$ are the momentum of entrance and exit channels, respectively; and $E$  is collision energy of entrance (m) channel. $V_m$ and $V_n$ are the potential energies of the entrance ($m$) and exit ($n$) channels, respectively. $\psi_\ell^i(k_ir)$ is the wave function of  $\ell$-th partial wave for $i$-th channel of momenum $k_i$. The total radiative transfer \cite{Zygelman1} from the upper state ($m$) to the lower state ($n$) is given by 
\bea 
\sigma_{rt} = \frac{\pi}{k_m^2}\sum_\ell^\infty ( 2\ell+1) \left [ 1-\exp(-4\zeta_{\ell})\right ] \label{rt}
\eea
where
\bea
\zeta_{\ell} = \frac{\pi}{2}\int_0^\infty |\psi_\ell^m(k_mr)|^2 A_{nm}(r) dr
\eea
 is a phase shift and 
\bea
A_{nm}(r) =\frac{4}{3} D^2(r)\frac{|V_n(r) - V_m(r)|^3}{c^3}
\eea
is the transition probability.
\begin{figure}
\includegraphics[height=2.0in,width=\columnwidth]{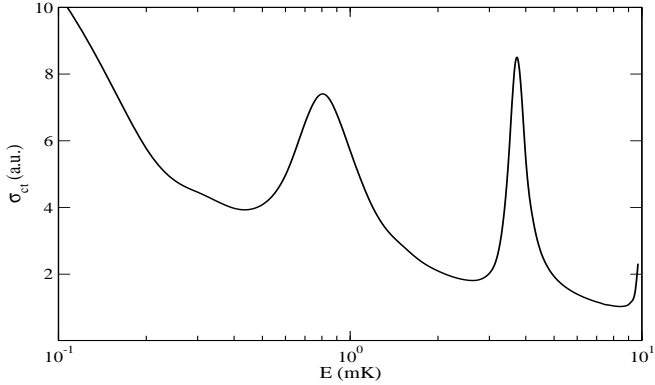}
\caption{Charge transfer scattering crosssection $\sigma_{ct}$ (in a.u.) of (LiBe)$^+$ system is plotted against collisional energy $E$ (in mK).}
\end{figure}

The ground continuum atom-ion pair, formed by radiative charge transfer process, can be photoassociated to  form excited molecular ion. This process is basically one photon PA process. The photoassociation rate coefficient is given by 
\bea
K_{PA} = \left \langle \frac{\pi v_{r}}{k^2}\sum_{\ell=0}^\infty (2\ell+1) |S_{PA}(E,\ell,w_L)|^2 \right  \rangle \label{kpa}
\eea
where $v_{r} = \hbar k/\mu$ is the relative velocity of the two particles and $\langle \cdots\rangle$ implies averaging over thermal velcity distribution. Here $S_{PA}$ is  S matrix element given by 
\bea
|S_{PA}|^2 = \frac{\gamma \Gamma_{\ell} }{\delta_E^2 + (\Gamma_\ell + \gamma)^2/4}
\eea
where $\delta_E = E/\hbar +\delta_{vJ}$, $ \delta_{vJ} = \omega_L -\omega_{vJ}$ with $ E_{vJ} = \hbar\omega_{vJ}$ being binding energy of the excited ro-vibrational state, $\omega_L$ being the laser frequency and $\gamma$  the spontaneous line width. Thus PA rate is primarily determined by partial wave stimulated line width $\Gamma_\ell$  given by
\bea
\hbar\Gamma_{\ell} =\frac{8\pi^2 I}{3\epsilon_0 c}  h(J,\ell) |D_{vJ,l}|^2
\eea 
where 
\bea
D_{vJ,l} =\langle \phi_{vJ} \mid D(r) \mid \psi_{\ell} (kr) \rangle
\eea
is the radial transition dipole matrix element between the continuum and  bound state wave functions $\psi_{\ell} (kr)$ and $\phi_{vJ} (r)$, respectively. $I$ is the intensity of laser, $c$ is the speed of light and $\epsilon_0$ is the vacuum permittivity. Here $h(J,\ell)$ is H\"{o}nl London factor \cite{honl} which in the present context is given by
\bea
 h(J,\ell) = (1 + \delta_{\Lambda^{\prime}0} + \delta_{\Lambda^{\prime\prime}0} -2\delta_{\Lambda^{\prime}0}\delta_{\Lambda^{\prime\prime}0})\nonumber\\
(2J+1)(2\ell+1) \left(\begin{array}{ccc}
 J&1&\ell\\
-\Lambda^{\prime}&\Lambda^{\prime}-\Lambda^{\prime\prime} &\Lambda^{\prime\prime}
\end{array}\right)^2
\eea
where $\Lambda^{\prime}$ and $\Lambda^{\prime\prime}$ are the projections of the total electronic orbital angular momentum of the excited and ground states, respectively, on molecular axis and 
$(\cdots)$ is the Wigner 3j symbol. 
The spontaneous line width $\gamma$ of the excited state $(v,J)$ is given by
\bea
\hbar\gamma  &=& \frac{1}{3\pi\epsilon_0  c^3}  \left [ \int (\Delta E)^3|\langle \phi_{vJ} \mid D(r) \mid \psi_E \rangle|^2 dE  \right. \nonumber \\
&+& \left. \sum_{v',J'} \Delta_{v'J'}^3 |\langle \phi_{vJ} \mid D(r) \mid \phi_{v'J'} \rangle|^2 \right ]
\label{spline} \eea 
where $\Delta E =(E_{vJ}-E)/\hbar$,  $\Delta_{v'J'} =(E_{vJ}-E_{v',J'})/\hbar $, $\psi_E$ is the scattering wave function and $\mid \phi_{v'J'} \rangle $ stands for all the final bound states to which the excited state can decay spontaneously.

\begin{figure}
\includegraphics[height=2.0in,width=\columnwidth]{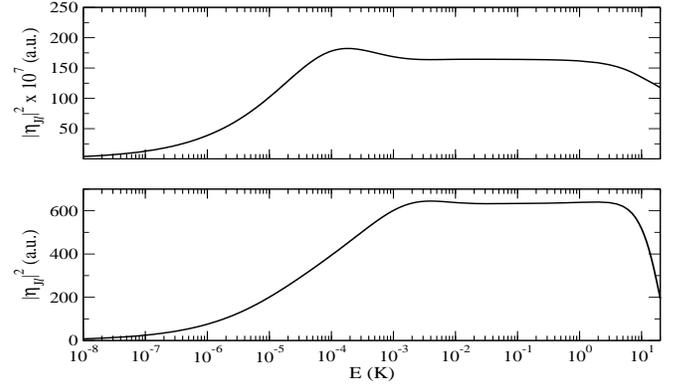}
\caption{ Square of Franck Condon overlap integral $|\eta_{J\ell}|^2$  (in a.u.) for Li-Li$^+$ (upper) and (LiBe)$^+$ (lower) is plotted against $E$ (in K). In the upper panel, $|\eta_{J\ell}|^2$ is mutiplied by a factor of $10^7$.}
\end{figure}

%============================================================================================================================
\section{Results and Discussion}

Standard renormalized Numerov-Cooley method \cite{Johnson} is used to calculate the bound  and scattering state wave functions. The molecular transition dipole matrix element of (LiBe)$^+$ system is calculated using GAMESS. This matrix element strongly depends upon separation and goes to zero at a large $r$ as shown in Fig.3.  In Figs. 4 and 5, we have plotted the excited and ground state  elastic scattering cross section $\sigma_{el}$ as a function of energy $E$ for Li + Be$^{+}$ and Li$^{+}$ + Be collisions, respectively.  We find that at least 35 partial waves are required to get converging results on elastic scattering for energies higher than 1 $\mu$K. In our calculations we have used 51 partial waves. At high energies, for both the cases, $\sigma_{el}$ decreases as $E^{-\frac{1}{3}}$. The proportionality constant $c$ in the expression $\sigma_{el} (E \rightarrow \infty) = c E^{-\frac{1}{3}}$ calculated using Eq. (\ref{sig}) for excited $2^1\Sigma^+$ and ground $1^1\Sigma^+$ potentials are 2936 and 1091 a.u., respectively, whereas linear fit to $\sigma_{el}$ vs. $E$ curves provides $c =$ 3548 and 1335 a.u., respectively. Figures 6 and 7 exhibit  s- and d-wave partial scattering cross section as a function of energy for Li$^+$+Be and Li+Li$^+$ collisions, respectively.  These figures show that the Wigner threshold behavior begins  to set in as the collision energy decreases below 0.1 $\mu$K. In Fig. 8,  we have plotted total elastic scattering cross section for Li+Li$^+$ collisions in  $^2\Sigma_g^+$  and $^2\Sigma_u^+$ potentials. 

Starting from the low energy continuum state of Li $+$ Be$^+$ collision in the  $2^1\Sigma^+$ potential, there arise two possible radiative transitions by which the system can go to the ground electronic state  1$^1\Sigma^+$. One is continuum-continuum and the  other is continuum-bound dipole transition. The transition dipole moment as a function of separation as shown in Fig. 3 shows that  the dipole transition probability will vanish as the separation  increases above 20$a_0$. So, a dipole trasition has to take place at short separations.
\begin{figure}
\includegraphics[height=2.5in,width=\columnwidth]{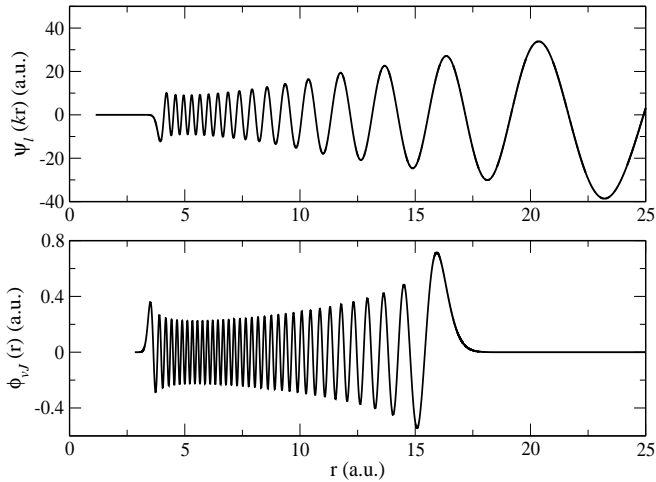}
\caption{ Energy-normalized s-wave ground scattering (upper) and unit normalized excited bound (lower) wave functions of (LiBe)$^+$ system are plotted as a function of separation $r$. }
\end{figure}
Let us consider radiative transfer processes from the upper (2$ ^1\Sigma^+$) to the lower (1$ ^1\Sigma^+$)
state of (LiBe)$^+$. We then need to apply the formulae (\ref{ct}) and (\ref{rt}) where $m \equiv 2 ^1\Sigma^+$ and $n \equiv 1^1\Sigma^+$ in our case. Continuum-continuum charge transfer cross section $\sigma_{ct}$ between 2$ ^1\Sigma^+$ and 1$ ^1\Sigma^+$ states of (LiBe)$^+$ system is plotted against $E$ in Fig.9.  We evaluate the photoassociative (continuum-bound) transfer cross section by subtracting $\sigma_{ct}$ from the total radiative transfer cross section $\sigma_{rt}$ calculated using the formula (\ref{rt}).    At energy $E=0.1$ mK, $\sigma_{ct}$ and the photoassociative transfer cross section are found to be  10.39 a.u. and 0.03 a.u., respectively.  Thus we infer that the continuum-continuum radiative charge transfer process dominates over the radiative association process. Also, we notice that  $\sigma_{ct}$ is smaller than both the excited and ground state elastic scattering cross sections $\sigma_{el}$ (as given in Fig.4 and 5, respectively) by  several orders of magnitude. 
\begin{figure}
\includegraphics[height=2.5in,width=\columnwidth]{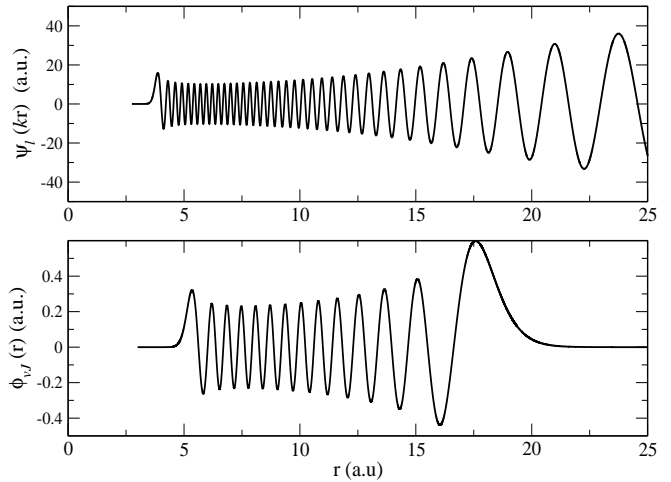}
\caption{ Same as in Fig.11 but for Li-Li$^+$ system.}
\end{figure}

Molecular dipole transitions between two ro-vibrational states or between continuum  and bound states are governed by Franck-Condon principle. According to this principle, for excited vibrational (bound) states, bound-bound or continuum-bound transitions primarily occur near the  turning points of bound states. In general,  highly excited vibrational state wave functions of diatomic molecules or molecular ions  have their maximum amplitude near the outer turning  points. Spectral intensity is proportional to the  overlap integral.This means that the  spectral intensity for a continuum-bound transition would be significant  when the continuum state has a prominent node near the outer turning point of the bound state.  For transitions between two highly excited bound states, Franck-Condon principle implies that the probability of such transitions would be significant when the outer turning points of these two bound states lie nearly at the same separation. The upper panel of Fig.10 shows the variation of the square of franck Condon overlap integral $|\eta_{J\ell}|^2$ between the ground s-wave ($\ell =0$) scattering  and the excited ro-vibrational ($v=26, J=1/2$) states of Li-Li$^+$ system as a function  the collision energy $E$. The lower panel of Fig.10 displays the same as in the upper panel but for (LiBe)$^+$ system with $v=68$ and $J=1$. The excited ro-vibrational state $v=26, J=1/2$ of Li-Li$^+$ is very close to dissociation threshold while the excited ro-vibrational state $v=68, J=1$ of (LiBe)$^+$ system is a deeper bound state. Thses two excited states are so chosen such that free-bound Franck-Condon overlap integral for both the systems become significant. Comparing these two plots, we find that  $|\eta_{J\ell}|^2$ of Li-Li$^+$ system is smaller than that of (LiBe)$^+$ system by seven orders of magnitude.  To understand why the values $|\eta_{J\ell}|^2$ for the two systems are so  different, we plot the the energy-normalized s-wave ground scattering and the bound state wave functions of (LiBe)$^+$  system in Fig.11  and  those of Li-Li$^+$ system in Fig.12. A comparison of  Figs.11 and 12 reveals that,  while in the case of  (LiBe)$^+$ the maximum  of the excited bound state wave function  near the outer turning point coincides nearly with a prominent antinode of the scattering wave fucntion, in the case of 
Li-Li$^+$  the maximum of the bound state wave function near the outer turning point almost coincides with a minimum (node) of the scattering wave function. These results indicate that  the possibility of the formation of excited LiLi$^+$ molecular ion via PA is much smaller than that of (LiBe)$^+$ ion. We  henceforth concentrate on PA of (LiBe)$^+$ system only.

\begin{figure}
\includegraphics[height=2.5in,width=\columnwidth]{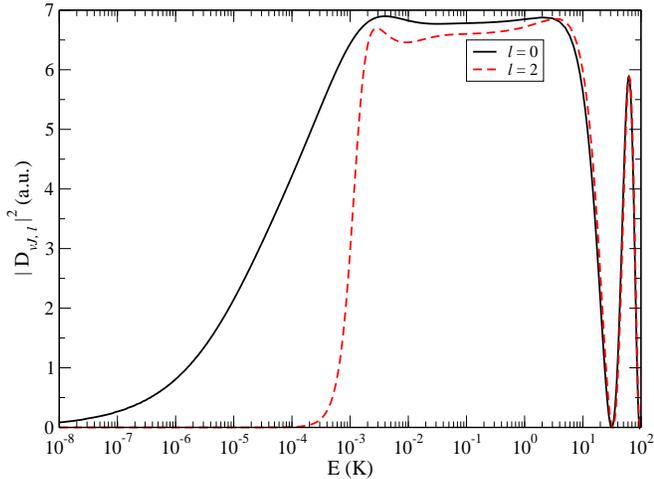}
\caption{(Color online) Square of free-bound radial transition dipole moment ($|D_{vJ,l}|^2$) (in a.u.) for ground continuum states with  $\ell = 0$ (solid) and $\ell = 2$ (dashed) and excited bound ro-vibrational level with ${\it v} = 68$ and $J = $ 1}
\end{figure}
\begin{table}
\caption{Ro-vibrational energy ($E_{{\it v}J}$), inner ($r_i$) and outer turning points ($r_o$) of two selected bound states of (LiBe)$^+$ molecular ion - one bound state in excited ($2^{1}\Sigma^+$) and the other in ground ($1^{1}\Sigma^+ $) potential. The energy $E_{{\it v}J}$ is measured from the threshold of the respective potential.}
\begin{tabular}{c  c  c  c  c  c  c c c c c}
\hline 
\multicolumn{1}{c}{{Potential}} & \multicolumn{1}{c}{\vline} & \multicolumn{1}{c} {$v$} &  \multicolumn{1}{c}{\vline} & \multicolumn{1}{c} {$J$} &  \multicolumn{1}{c}{\vline}  & \multicolumn{1}{c} {$E_{vJ}$ (a.u.)} &  \multicolumn{1}{c}{\vline}  & \multicolumn{1}{c} {$r_i$} (a.u.) &  \multicolumn{1}{c}{\vline} &\multicolumn{1}{c}{$r_o$ (a.u.)}  \\
\hline
 $2^{1}\Sigma^+$ & \vline & 68 &\vline & 1 & \vline &  -3.30$\times 10^{-3}$ & \vline & 3.4 &  \vline & 16.3 \\
$1^{1}\Sigma^+$ & \vline & 29 & \vline  & 0 &  \vline &  -0.25$\times 10^{-3}$  &  \vline &  3.8 & \vline & 16.6  \\
\hline
  \end{tabular}
\label{tb4}
\end{table} 
We next  explore the possibility of PA in Li$^+$-Be cold collision in the presence of laser light.  As discussed before, continuum-bound  molecular dipole transition matrix element depends on the degree of overlap between continuum and bound states.  PA rate (\ref{kpa}) is proportional to  the square of free-bound radial transition dipole moment element $|D_{vJ,l}|^2$.  In Fig. 13 we plot $|D_{vJ,l}|^2$ against $E$  for s- ($\ell = 0$)  and d-wave ($\ell = 2$) ground scattering states  and  $v = 68$ , $J = 1$ excited molecular state. It is clear from this figure that the contributions of both $\ell =$ 0 and $\ell =$ 2 partial waves are comparable above enegy corresponding to 0.1 mK. At lower energy ($E<0.1$ mK), only s-wave makes finite contribution to the the dipole transition.  Figure 14 exhibits $|D_{vJ,l}|^2$ as a fucntion of $E$  for the transition from s-wave ($\ell = 0$) scattering state of the excited ($2^1\Sigma^+$) continuum  to the ground ($1^1\Sigma^+$) ro-vibrational state with $v = 36$ , $J = 1$. A comparison between the Figs.13 and 14 reveals that the probability for the transition from the upper continuum to the ground bound state is smaller  by several orders of magnitude than that from ground continuum to an excited bound state.
In Fig 15, we have plotted the rate of photoassociation $K_{PA}$ as a function of temperature $T$ for laser frequency tuned at PA resonance. The ion-atom PA rate as depicted in Fig.15 is comparable to the typical values of rate of neutral atom-atom PA at low laser intesities. In Fig. 16 we have plotted the rate of photoassociation as a function of laser intensity at a fixed temperature $T = 0.1$ mK to show the saturation effect that occurs around intensity $I = 50$ kW/cm$^2$. Thus the formation of excited (LiBe)$^{+}$ molecular ion by photoassociating colliding  Li$^+$ with Be with a laser of moderate intensity appears to be a feasible process. 
\begin{figure}
\includegraphics[height=2.25in,width=\columnwidth]{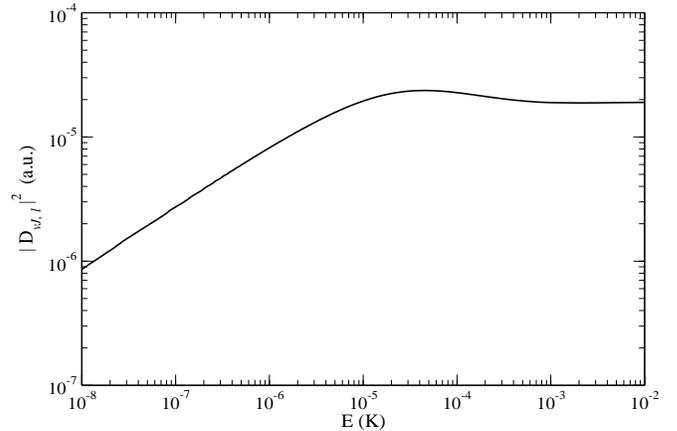}
\caption{Same as in Fig.13  but for excited continuum state with  $\ell = 0$ and ground ro-vibrational state with $ v=36$ and $J=1$. }
\end{figure}

\begin{figure}
\includegraphics[height=1.95in,width=2.5in]{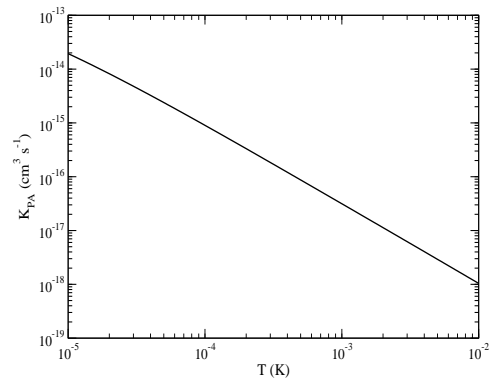}
\caption{Rate of photoassociation $K_{PA}$ (in cm$^3$ s$^{-1}$)  of (LiBe)$^+$ is plotted against temperature (in K) at $I=1$ W/cm$^2$ and $\delta_{vJ} = \omega_L -\omega_{vJ} = 0$}
\end{figure}

Now we discuss the possibility of formation of ground state molecular ion by stimulated Raman-type process by applying a second laser tuned near a bound-bound transition between the excited and ground potentials. Let us consider two selected bound states whose salient features are given in Table-II. 
The outer turning points of these two bound states almost coincide implying the existence of a large Franck-Condon overlap between them. 
To see whether coherent laser coupling between these two bound states is possible or not, we calculate 
Rabi frequency $\Omega$ given by 
\bea
\hbar\Omega = \left(\frac{ I}{4 \pi c \epsilon_0}\right)^{\frac{1}{2}} |\langle v,J \mid \vec{D}(r).\hat{\epsilon}_L \mid v^{\prime},J^{\prime} \rangle| 
\eea 
where $\hat{\epsilon}_L $ is the unit vecot of laser polarization and 
$\mid v, J \rangle$ and $\mid v^{\prime},J^{\prime} \rangle$ are the two bound states with $\langle r \mid v,J \rangle = \phi_{vJ}(r)$. Rabi frequency corresponding to this bound-bound  transition is found to be 285 MHz for laser intensity $I = 1 $ kW/cm$^{-2}$. Comparing this value with the spontaneous line width $\gamma = 57$ kHz of the excited bound state calculated using the formula (\ref{spline}), we infer that even at a low laser intensity which is far below the saturation limit, bound-bound Rabi frequency $\Omega$ exceeds  $\gamma$ by several orders of magnitude. This indicates that it may be possible to form ground molecular ion by stimulated Raman-type process with two lasers.

\begin{figure}
\includegraphics[height=2.5in,width=\columnwidth]{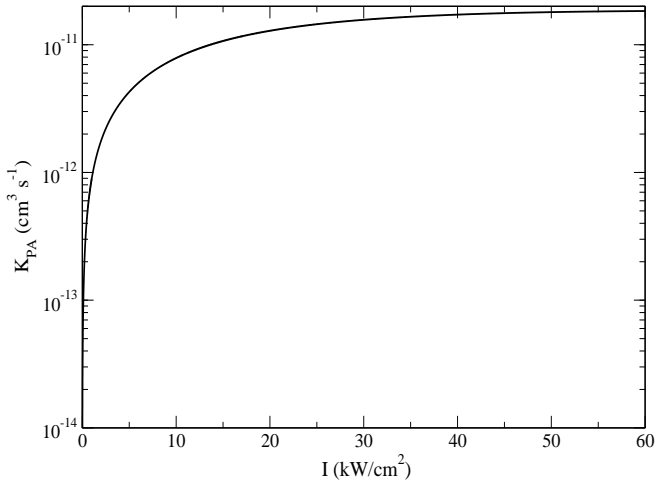}
\caption{ $K_{PA}$ (in cm$^3$ s$^{-1}$)  of (LiBe)$^+$ is plotted as a function of laser intensity $I$ (in kW/cm$^2$) at temperature $T = 0.1$ mK with laser tuned at PA resonance.}
\end{figure}

\section{CONCLUSION}
In conclusion, we have shown that alkaline earth metal ions  immersed in Bose-Einstein condensates of alkali atoms can give rise to a variety of cold chemical reactions. We have analyzed in detail the elastic and inelastic processes that can occur in a system of a Beryllium ion interacting with cold Lithium atoms. We have predicted the formation of translationally and rotationally cold
(LiBe)$^+$ molecular ion by photoassociation. Theoretical understanding of low energy atom-ion scattering and reactions may be important for probing dynamics of quantum gases. Since both Bose-Einstein condensation and fermionc superfluidity have been realized in atomic gases of Lithium, understanding cold collisions between Lithium and Beryllium ion may be helpful in probing both bosonic and fermionic superfluidity. In particular, this may serve as an important precursor for generating and probing vortex ring in Lithium quantum gases.

\section*{Acknowledgments}
AR is grateful  to CSIR, Government of India, for a support. We are thankful to P. Ghosh, Presidency College, Kolkata for his help in computation.

%-------------------------------------------------------------------------------------------------------------------

\end{document}